%% file: paper6.tex
\documentstyle[]{l-aa}
\include{eps}

\begin{document}
\thesaurus{}
\title{SN\,1995ah-the first supernova observed in a Blue Compact Dwarf galaxy
\thanks{Partly based on observations obtained at the European Southern
Observatory, La Silla, Chile and at the German-Spanish Observatory, Calar
Alto, Almeria, Spain} }
\author{Cristina C. Popescu \inst{1,5}\thanks {Present address: MPIK,
Saupfercheckweg1,  D-69117 Heidelberg, Germany}
\and Piero Rafanelli \inst2
\and Stefano Benetti \inst3
\and  Ulrich Hopp \inst{4}\thanks{Visiting astronomer at Calar Alto}
\and Kurt Birkle          \inst1
\and  Hans Els\"asser \inst1
}

\offprints{Cristina C. Popescu}
\institute{Max Planck Institut f\"ur Astronomie, K\"onigstuhl 17, 
           D--69117 Heidelberg, Germany
\and Department of Astronomy, University of Padova, Vicolo Osservatorio 5,
           35122 Padova, Italy
\and European Southern Observatory, Alonso de Cordova 3107, Vitacura,
     Casilla 19001 Santiago 19, Chile 
\and Universit\"atssternwarte M\"unchen, Scheiner Str.1, 
           D--81679 M\"unchen, Germany   
\and The Astronomical Institute of the Romanian Academy, Str. Cu\c titul de
           Argint 5, 75212, Bucharest, Romania}
\date{Received 02.06.1997; Accepted 21.07.1997}
\maketitle

\begin{abstract}
We present the properties of the supernova SN\,1995ah discovered in a Blue
Compact Dwarf galaxy (BCD) around 10 days after the maximum. This is the first 
supernova event observed in a BCD. The photometric and 
spectroscopic data suggest that SN\,1995ah  is a Type II supernova and
could belong to the rare Bright SNII Linear subclass, for which $\langle 
{\rm M}_{R} \rangle \sim \langle {\rm M}_{B} \rangle  = -18.9 \pm 0.6$ at 
maximum light. 
\end{abstract}

\section{Introduction}

The Type II supernova 1995ah was discovered on February 2, 1995 by 
Popescu et al. (1995) during the follow-up spectroscopy of a survey for 
emission-line galaxies (Popescu et al. 1996a); the discovery of this supernova
 was therefore serendipitous. The survey itself was conducted 
with the aim to search for emission-line galaxies towards nearby voids 
(Popescu et al. 1996b, Popescu et al. 1997) and to study the spatial
distribution of the dwarf 
emission-line galaxies in comparison with the population of normal (giant) 
galaxies. The candidates for emission-line objects were selected from the 
objective-prism plates of the Hamburg Quasar Survey (HQS) (Hagen et al. 1995). 
The selection of 
candidates was done on digitised spectra and was based on the presence of the 
[OIII] ${\lambda}$5007 line. 

The galaxy HS0016+1449 was selected as a candidate in
the frame of the above mentioned survey and was confirmed as a galaxy with
emission-lines during one of the follow-up spectroscopic runs at the 2.2\,m 
telescope at Calar Alto. The galaxy was previously unknown in literature, and 
its main characteristics are given by Popescu et al. (1996a) (Tables 4, 5). 
The spectrogram together with some further direct images of the galaxy
indicated that HS0016+1449 is a typical Blue Compact Dwarf galaxy with
 high degree of ionization and blue continuum, at a redshift z=0.0147.  A 
later spectrogram of the galaxy led to the discovery of the supernova 1995ah. 
This is the first case of a supernova discovered in a Blue Compact Dwarf.

The compact galaxies were discovered and classified as a separate class of
objects  by Haro (1956) and Zwicky (1964), based on their high surface 
brightness appearance on the photographic plates. The whole family was later 
divided into two subclasses, red and blue, according to their integrated 
colours. The Blue Compact Dwarfs received 
considerable attention when it was realised that these objects are places where 
strong star formation is occuring (Searle \& Sargent 1972). Late type 
spirals and 
irregular galaxies have been long known as centres of active star formation with
 their appearance in the blue region of the spectrum largely determined by the
 emission from young stars. However, in late 
type spirals and as well as in irregulars the bulk of the stellar mass lies 
not in the young
 stars but in an older stellar population which contributes most of the 
emission in the far red and infra-red spectral regions. Sargent \& Searle 
(1970) discovered that among the BCDs there are members in which the young
 stellar population dominates in blue and in the far infrared. It was
proposed that the BCDs are galaxies undergoing a strong burst of
star formation and it may even be that they produce primarily 
massive stars rather than stars with a 
luminosity function similar to that of our solar neighbourhood. Sargent \& 
Searle termed these objects \lq\lq isolated extra-galactic HII regions\rq\rq,
but nowadays the most commonly denomination is \lq\lq HII galaxies\rq\rq.

While the name of BCDs was mainly used for objects that
 were classified on morphological criteria (Binggeli et al. 1985 - for the 
Virgo Cluster Catalog), the term of HII galaxies was introduced for objects 
discovered on spectroscopic surveys for emission-line objects. The class of 
BCDs contains however a rather heterogenous family of objects, with different 
properties (Campos-Aguilar et al. 1993).


Typical diameters, intrinsic luminosities and H${\alpha}$ luminosities
of the HII galaxies range between $10\,{\rm kpc} > {\rm D} > 2\,{\rm kpc}$, 
$-18.5 < {\rm M}_B < -14$, and 
$3\times 10^{41} > {\rm L}({\rm H}{\alpha}) > 10^{35}$\,erg/sec, respectively.

\begin{table}[htp]
\caption[]{Dates and set-up of the spectroscopic observations: CA: Calar Alto, 
ESO: European Southern Observatory, BCCS: Boller \& Chivens Cassegrain 
Spectrograph, EFOSC1: ESO Faint Object Spectrograph and Camera 1. }
\begin{tabular}{llllll}
                      & & &\\
\hline\hline
                 & & &\\
Date (UT)        & Oct., 8.141 & Feb., 2.792 & Oct., 14.211\\
                 & 1994        & 1995        & 1995\\
Telescope        & 2.2\,m, CA & 2.2\,m, CA  & 3.6\,m, ESO\\
Instrument       & BCCS & BCCS & EFOSC1\\
Detector         & TEK13 & TEK6 & TEK26\\
Pixel Size ($\mu$)  & 24 & 24 & 27\\
Slit width (${\prime\prime}$) & 2 & 2 & 2\\
P.A.                  & 90. & 90. & 216.\\
Pixel number          & 1024$\times$1024 & 1024$\times$1024 & 512$\times$512\\
Dispersion (\AA/pixel) & 5.1 & 5.2 & 6.3 \\
Resolution (\AA) & 12 & 12 & 18 \\
Spectral Range (\AA)  & 3600-9000 & 3460-8600 & 3700-6800\\
Exposure time (s)     & 600 & 459 & 1800\\
& & &\\
\hline
\end{tabular}
\end{table}

\begin{table}[htp]
\caption[]{Dates and set-up of the photometric observations:  CA: Calar Alto, 
 ESO: European Southern Observatory, CAFOS: Calar Alto Faint Object 
Spectograph, PFFR: Prime Focus Focal Reducer + 2 lens corrector.}
\begin{tabular}{llllll}
                      & & &\\
\hline\hline
                  & & &\\
Date (UT)         & Feb., 03.786 & Oct., 14.197 & Oct., 17.935\\
                  & 1995         & 1995         & 1995\\
Telescope                          & 2.2\,m, CA  & 3.6\,m, ESO & 3.5\,m, CA \\
Instrument                         & CAFOS  & EFOSC1 & PFFR\\
Detector                           & TEK13  & TEK26 & TEK7\\
Pixel Size ($\mu$)                 & 24  & 27 & 24\\
Pixel number        & 1024$\times$1024  & 512$\times$512 & 1024$\times$1024\\
Pixel scale ($^{\prime\prime}$/pixel) & 0.511  & 0.610 & 0.406\\
Filter                             & R  & R & 703/34\,nm\\
Exposure time (s)                  & 60  & 120 & 600\\
& & &\\
\hline
\end{tabular}
\end{table}

The paper is organised as follows. In $\S$ 2 we describe the spectroscopic and
the photometric properties of the supernova and of the host galaxy, in 
 $\S$ 3 we make some extimates on the SNR of 
this kind of galaxies and $\S$ 4 contains the summary.

\section{The observations}

Table 1 gives the set-up of the spectroscopic observations of the galaxy
HS0016+1449 before the supernova exploded, when it was close to 
maximum and 254 days after the discovery. The first two observations were taken
 with the 2.2\,m telescope at the German-Spanish Observatory at Calar Alto 
(Almeria, Spain) and the last observation was obtained with the ESO 3.6\,m
telescope. The spectroscopic data were reduced using both the MIDAS and IRAF 
packages. In all cases a standard procedure was applied, namely: bias 
subtraction, flat-field correction, extraction of the
1-dimensional spectra from the 2-dimensional one, transformation to the true
wavelength scale, correction for the atmospheric extinction using mean
extinction values for Calar Alto and La Silla, respectively, and
convertion to true flux from observations of spectrophotometric  standard 
stars.

The journal of the photometric observations is given in Table 2. The first
image was taken during the supernova explosion with the CAFOS (The Calar
Alto Faint Object Spectograph) at the 2.2\,m telescope at Calar Alto. A 
second R image was taken at ESO 253 days after the discovery of the supernova 
together with the spectroscopic 
observations described above. The deepest direct image was taken at the Prime 
Focus of the 3.5\,m Focal Reducer at Calar Alto. 
 We also observed several open-clusters containing photometric standard stars, 
namely NGC2264 and NGC2419 at Calar Alto
and Ru149 at La Silla. All the direct images were taken in photometric
conditions. The photometry of the frames was performed using aperture 
photometry and the
software available in MIDAS. For the calibration of the frames we used the data
from Christian et al. (1985)(NGC2264, NGC2419) and Landolt (1992) (Ru149).
\begin{figure}[hp]
\plotfiddle{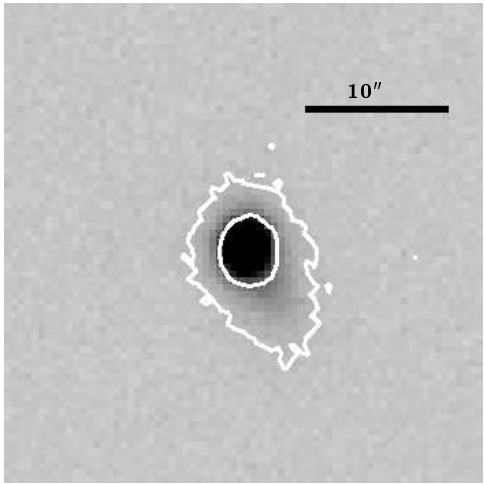}{3.0in}{0.}{40.}{40.}{-110}{-75}
\caption[]{ The R image of the galaxy HS0016+1449 during the supernova 1995ah
explosion (February 3, 1995). The direct image shows the supernova (traced by
the inner contour plot) which 
outshines the diffuse emission from the host galaxy (traced by the outer
contour plot).  North is up and east is left.}
\end{figure}

\begin{figure}[hp]
\plotfiddle{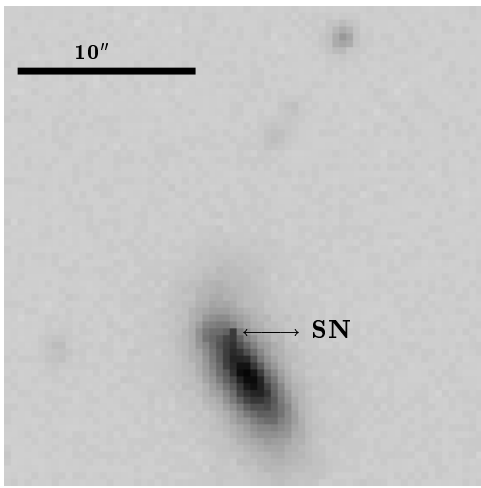}{3.0in}{0.}{40.}{40.}{-110}{-60}
\caption[]{A deep 703/34\,nm image of the galaxy HS0016+1449 256 days after 
the first exposure. The direct image shows that the supernova is still 
visible on the Northern-Eastern edge of the galaxy. North is up and east is 
left. }
\end{figure}

\begin{figure}[htb]
\plotfiddle{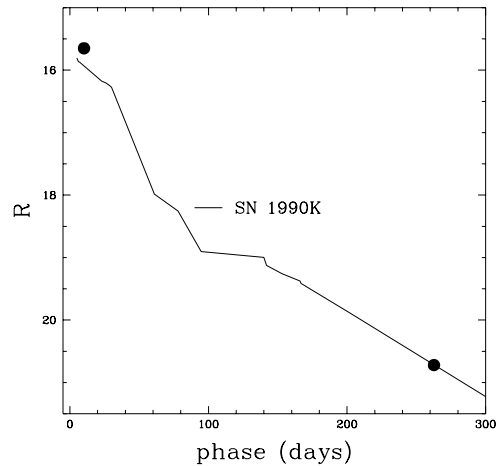}{3.0in}{0.}{35.}{35.}{-110}{-60}
\caption[]{Comparison of our data (solid dots) with SN\,1990K R light
curve (Cappellaro et al. 1995). The SN\,1990K light curve has been scaled to
SN\,1995ah distance and reddening, assuming an ${\rm H}_{0}=75\,{\rm km}\,
{\rm s}^{-1}\,{\rm Mpc}^{-1}$.}
\end{figure}

\subsection{The photometry}

The first direct frame taken on February 3, 1995 (Fig. 1) is dominated by
the supernova image which outshines the faint underlying galaxy. 
A deep 600\,s exposure taken 256 days later (October 17, 1995) (Fig. 2) shows 
the supernova which faded substantially. The remnants
of the supernova are still visible on the northern-eastern edge of the galaxy. 
The same features are present on the red ESO image
taken on October 14, 1995.  The distance of the supernova from the centre of
the galaxy is 2.57$^{\prime\prime}$, at a PA=18.4$^{\circ}$. At a redshift
of z=0.0147, this corresponds to 0.7\,kpc. Throughout this paper a
Hubble constant of 75\,km/s/Mpc is considered. The galaxy itself
shows little structure; a hot-spot at its center is embedded in a faint and
structureless envelope. Some interesting features seem to surround the galaxy,
maybe nearby companions and remnants of tidal interactions. It was suggested 
(Taylor et al. 1994, Taylor 1997) that interaction with nearby companions may
play an important role in triggering bursts of star formation in HII
galaxies. However, the faint features that seem to surround the galaxy could
result from projection effects and further observations would be required to
confirm the real nature of the \lq\lq companions\rq\rq.
The diameters of
the galaxy measured along the major and the minor axis were
D=14.5$^{\prime\prime}$ (4.2\,kpc) and d=6.8$^{\prime\prime}$ (1.9\,kpc), 
respectively.

From our photometric data we computed the apparent magnitudes of the SN and of
the host galaxy. For the galaxy HS0016+1449 we found an apparent magnitude 
of R=17.7 which gives an absolute magnitude of M$_{R}=-16.3$, after
correction for a galactic extinction of A$_{R}=0.09$ (taken from Burstein \&
Heiles 1984). We estimate that the errors in our magnitude determinations are
around 0.2\,mag. The 
supernova was initially as bright as 
R=15.8 (February 3) and it faded down to R=20.7 (October 14), 
253 days later. The corresponding absolute magnitudes of the supernova were
 M$_{R}=-18.2$ (February 3) and M$_{R}=-13.2$ (October 14). The first data
point we believe to be measured 10 days after the maximum, as our spectral
characteristics indicates (see next subsection). Given the magnitude
of the supernova near maximum we suggest that SN\,1995ah could belong
to the rare Bright SNII Linear subclass (Patat et al. 1994), 
for which $\langle {\rm M}_{R} \rangle \sim \langle {\rm M}_{B}
\rangle  = -18.9 \pm 0.6$ at maximum light. We made then an attempt to 
compare our two
photometric measurements with the light curves of another Bright
SNIIL, SN\,1990K (Cappellaro et al. 1995) (Fig. 3). The SN\,1990K light curve
 has been scaled to the distance and
reddening of SN\,1995ah. We assumed January 24th as the date of
maximum of the SN\,1995ah light curve. The agreement is quite good and
this tells us that the amount of energy involved in the SN\,1995ah and
SN\,1990K explosions could be similar.

\newpage
\subsection{The spectroscopy}

\begin{figure}[htb]
\plotfiddle{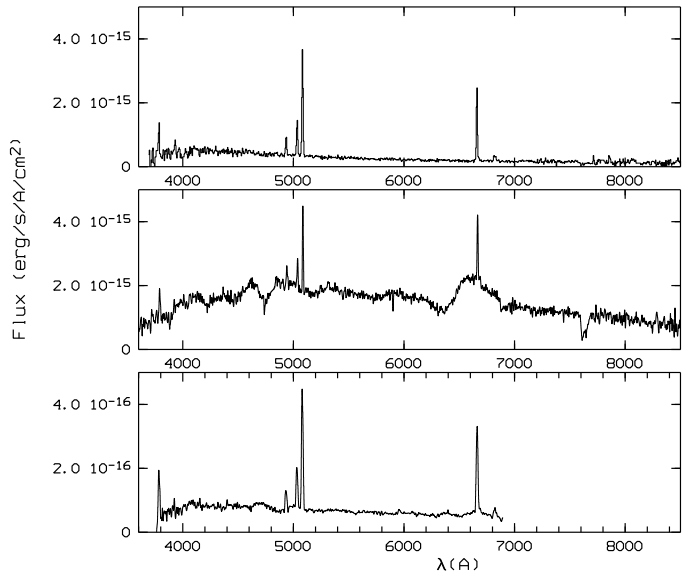}{4.0in}{-90.}{48.}{48.}{-205}{280}
\caption[]{The spectra of HS0016+1449. Top pannel: October 8, 1994; middle 
pannel: February 2, 1995; bottom pannel: October 14, 1995.}
\end{figure}

The top pannel of Fig. 4 gives the first spectogram of the galaxy taken on 
October 8, 1994.
The spectrum shows the typical narrow emission-lines of a high-ionization 
HII galaxy, namely [OII] ${\lambda}{\lambda}$3726,3731,  
[NeIII] ${\lambda}$3869, H${\beta}$ ${\lambda}$4861,   
[OIII] ${\lambda}{\lambda}$4959,5007, 
H${\alpha}$ ${\lambda}$6563 and [SII] ${\lambda}{\lambda}$6716,6731.

A second spectrum of the galaxy taken on February 2, 1995 (Fig. 4, middle
pannel) is
dominated by the features of a Type II supernova around 10 days after the
maximum; the spectrum is characterised by broad emission-lines 
with P-Cyg absorption on which the narrow-emission lines of the galaxy are
superimposed. 

The third spectrum taken on October 14, 1995 (Fig. 4, bottom pannel) shows 
that traces of the supernova are still visible, especially in the broad 
emission of 
the H${\alpha}$.

We first analyse the spectroscopic properties of the host galaxy with the
aim of proofing the membership of HS0016+1449 to the class of HII galaxies. 
In Table 3 we give the line ratios of the bright narrow emission-lines 
with respect to the H${\beta}$ line. The line ratios are corrected for the
extinction in our own Galaxy as well as for the internal extinction in the
galaxy itself. The intrinsic ratios 
I(${\lambda}$)/I(H${\beta}$) were
calculated using equation (7.6) from Osterbrock (1989). The reddening 
coefficient c(H${\beta}$) was derived 
from the H${\alpha}$/H${\beta}$ decrement and the value of 
c(H${\beta}$) 
was computed assuming that the intrinsic Balmer-line ratios are equal to the 
case B recombination values from Osterbrock (1989), 
H${\alpha}$/H${\beta}$=2.87.
The values of the line ratios and of the c(H${\beta}$) were separately computed
for the spectrum taken on October 8, 1994  (Column A) and for the one 
taken on October 14, 1995 but for different regions of the galaxy (Columns B,
C, D, E, F). Because in the latter spectrum the slit was oriented along the 
major axis of the galaxy, we were able to extract separately the contribution
 of different regions, including the region of the supernova. Each region
was artificially extracted to contain the same number of pixels and corresponds
to 1.83$^{\prime\prime}$ (0.5\,kpc) width. The regions were
considered from NE to SW. Table 3
shows that the first spectrum has the same internal extinction as the regions
B, C and D (an average of c(H${\beta}$)=0.44) while for the regions E and F 
the extinction grows up to 1.92 and 2.04, respectively. The errors in the line
ratios are around $5\%$ for Column A, B, C and D and up to 30$\%$ for the 
regions E and F. We should notice that the latter errors are due to the bad 
S/N of the H${\beta}$. The corrected 
line ratios from the first spectrum (A) are approximately the same as 
those in the region C of the last spectrum, where the supernova features are 
still visible. Presumably the first spectrum observed with a standard BC 
Cassegrain
Spectograph was taken with the slit crossing the region C.  The spectrum of 
the region C contains also the lines of H${\gamma}$ and 
[OIII] ${\lambda}$4363, the latter being used to estimate the 
electronic temperature. 
The table also shows a tendency of decreasing the degree of the oxygen 
ionization, from region B to region F, from a 
([OII] ${\lambda}$3727)/([OIII] ${\lambda}$5007) ratio of 0.28 up to a ratio 
of 5.31.

\begin{table}[htp]
\caption[]{ Line ratios of the narrow emission-lines of the galaxy with respect
 to the H${\beta}$ line, corrected for the
extinction in our own Galaxy as well as for the internal extinction in the
galaxy itself. The values of the line ratios and of the c(H${\beta}$) are 
computed for the spectrum taken on October 8, 1994  (Column A) and 
for the spectrum taken on October 14, 1995 (Columns B, C, D, E, F).
}
\begin{tabular}{rllllll}
                       & & & & & & \\
\hline\hline
                       & & & & & & \\
                    & A & B & C & D & E & F\\
                      & & & & & & \\
\hline
                       & & & & & & \\
c(H${\beta}$)                          & 0.46  & 0.48 & 0.40 & 0.44 & 1.92 &
2.04 \\
                       & & & & & & \\
\lbrack OII\rbrack\, ${\lambda}$3727  & 1.85  & 1.16 & 1.61 & 5.08 & 37.64 &
98.29\\
H${\gamma}\, {\lambda}$4340           &       &      & 0.46 & & &\\
\lbrack OIII\rbrack\, ${\lambda}$4363 &       &      & 0.15 & & &\\
\lbrack OIII\rbrack\, ${\lambda}$4959 & 1.95   & 1.71 & 1.98 & 1.64 & 3.12 &
6.56\\
\lbrack OIII\rbrack\, ${\lambda}$5007 & 5.85   & 4.13 & 5.66 & 4.76 & 8.35 &
18.50\\
& & & & & & \\
\hline
\end{tabular}
\end{table}

\begin{figure*}[htb]
\plotfiddle{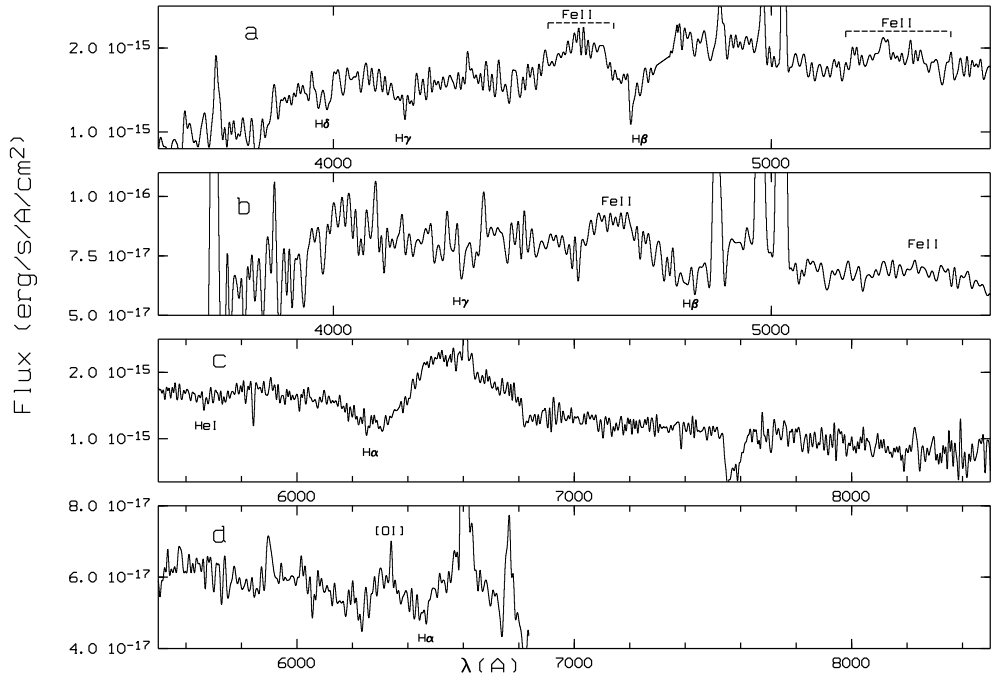}{4.0in}{-90.}{50.}{50.}{-185}{280}
\caption[]{Comparison between the supernova spectra taken shortly after the
maximum (February 1995)(panels a, c) and 254 days after (October 1995)(panels
b, d) . The upper panels (a, b) 
compare the blue part of the spectra and the lower panels (c, d) compare the 
red part of the spectra.  }
\end{figure*}

Though we divided our spectrum in different regions, mainly with the aim of
extracting the contribution of the supernova features, the spectrum is
characterised by two main spectral types, one of high ionization and low
reddening, corresponding to an extension of 5.49$^{\prime\prime}$ (1.5\,kpc) 
along the major axis of the galaxy (region B, C, D) and one of low ionization 
and high extinction, corresponding to an extension of 3.66$^{\prime\prime}$
(1.0\,kpc) (region E, F).

Using the
[OIII] ${\lambda}{\lambda}4959+5007$/[OII] ${\lambda}$4363 
ratio (for the region C) in the limit N$_{e}\rightarrow 0$
 (Osterbrock 1989, p. 121) we obtained a temperature of
T$_{e}=17500$\,K. From the ratio 
[SII] ${\lambda}$6716/[SII] ${\lambda}6713=1.17$  we 
determined an electron density N$_{e}=450\,$cm$^{-3}$ (see Osterbrock 1989, 
p. 134), assuming a temperature T$_{e}=10000\,$K. 
The N$_{e}$ determination can be recalculated using
our estimated temperature, which implies a 
correction of (10$^{4}$/T)$^{1/2}$ to the previous determination. The N$_{e}$
 becomes thus 350\,cm$^{-3}$. Using lower limit estimates for the 
[NII] ${\lambda}$6583 (again for the
high ionization region), and the diagnostic diagrams 
log([OIII] ${\lambda}$5007/H${\beta}$) vs 
log([NII] ${\lambda}$6583/H${\alpha}$) from Osterbrock (1989, 
p. 346) we found the location of our galaxy in the region occupied by the high 
ionization HII galaxies.  Using also the diagnostic diagrams 
log([OIII] ${\lambda}$5007/H${\beta}$) vs 
log([SII] ${\lambda}6716+6731$/H${\alpha}$) from 
Osterbrock (1989), p. 349, we found again that our galaxy is a typical HII
galaxy.

So both morphological and spectroscopic properties of the galaxy HS0016+1449
recommend it as a typical Blue Compact Dwarf.

\begin{table}[htp]
\caption[]{The expansion velocities (in km/s) in the supernova spectra}
\begin{tabular}{lll}
                       & & \\
\hline\hline
                       & & \\
line identification & \multicolumn{2}{c}{ velocity}\\
                      & &\\
\hline
                      & &\\
                       & February 2 & October 14\\
                 &      &\\
\hline
                 &      &\\
 H${\alpha}$     & 15900 & 6700\\
 HeI             & 12700 &     \\
 H${\beta}$      & 12200 & 4300\\
 H${\gamma}$     & 13200 & 4300\\
 H${\delta}$     & 10500 &     \\
 H${\epsilon}$   & 11400 &      \\
& \\
\hline
\end{tabular}
\end{table}

Next we will describe the spectroscopic features of the supernova. The first
spectrum of the supernova is typical of a SN Type II around 10 days 
after the maximum (Fig. 5\,a,c). It is characterised by broad 
emission lines with PCyg absorption on their blue side, of the Balmer series 
(visible up to H$_{\delta}$) and of the HeI ${\lambda}$5875 as well as by
 blends due to FeII (35), (37), (38), (42). The typical expansion velocities of the ejected shells measured 
from the positions of the
deepest features of the PCyg absorptions are given in Table 4. The upper limit
of the expansion velocity, derived from the wavelength of the blue wing of the
H$_{\alpha}$ absorption is ${\sim}$23000\,km/s. These high values
 confirm that SN\,1995ah has been
caught soon after explosion. As an example, SN\,1987A had a similar
expansion velocity only 6 days after explosion (Hanuschik et al. 1989). 

The second and last spectrum of the SN taken on October 14, 1995 (Fig. 5\,b,d) 
shows still traces
of the supernova continuum, of the broad Balmer emission-lines (visible up to
H${\gamma}$), of the FeII blends and [OI]. The expansion velocity as derived
from the FWHM of the emission lines, and given in Table 4, is a
factor $2\div3$  smaller than in the first spectrum, and similar to that of 
other SNII. No HeI ${\lambda}$5875 is more visible. 

\section{Supernova rates}

We can now relate the H${\alpha}$ luminosity of our galaxy to the SFR 
following Kennicutt
(1983). If we consider a Miller \& Scalo (1979) IMF:
\begin{eqnarray}
{\Phi}_{M} & = & {\Phi}_{0}\,{\rm}m^{-2.5} 
\end{eqnarray}
including stars in the $10\,{\rm M}_{\odot}\leq m\leq 100\,{\rm M}_{\odot}$ 
mass range, the transformation can be written as:
\begin{eqnarray}
{\rm SFR}\,(\geq 10\,{\rm M}_{\odot}) & = & \frac{{\rm L}({\rm H}{\alpha})}
{7.02\times 10^{41}{\rm ergs}\,{\rm s}^{-1}}{\rm M}_{\odot}\,{\rm yr}^{-1} 
\end{eqnarray}

Taking the H${\alpha}$ luminosity of 
HS0016+1449 L(H${\alpha}$)=$4.74\times 10^{39}\,$ergs/s, as measured in the 
spectrum of October
1995 we obtain a ${\rm SFR}\,(\geq 10\,{\rm M}_{\odot})=
0.68\times 10^{-2}\,{\rm M}_{\odot}\,{\rm yr}^{-1}$. 

The supernova rate SNR is related to the SFR through the following integrals:

\begin{eqnarray}
{\rm M}\,(\geq 10{\rm M}_{\odot}) & = & \int_{10}^{100}
{\rm m}\,{\Phi}_{0}\,{\rm m}^{-2.5}\,{\rm dm} = 0.71\,{\Phi}_{0}\,{\rm M}_{\odot}\\
{\rm N}\,(\geq 10{\rm M}_{\odot}) & = & \int_{10}^{100}
{\Phi}_{0}\,{\rm m}^{-2.5}\,{\rm dm} = 0.032\,{\Phi}_{0}
\end{eqnarray}

From (3) and (4) we obtain:
\begin{eqnarray}
{\rm SNR} & = & 0.045\,{\rm SFR}\,{\rm M}_{\odot}^{-1}
\end{eqnarray}

Thus, for a SFR of $0.68\times 10^{-2}\,{\rm M}_{\odot}\,{\rm yr}^{-1}$, the
corresponding supernova rate is $0.3\times 10^{-3}\,{\rm yr}^{-1}$, or 
${\rm SNR} = 0.58\pm 0.05\, {\rm SNu}$, where 
SNu=SN/100yr/10$^{10}\,{\rm L}_{\odot}({\rm B})$, M$_{\odot}({\rm B})=5.48$ and
we considered  B=17.6 for the galaxy (see Popescu et al. 1996a). The errors
were calculated from the uncertainties introduced by the H$\alpha$ flux
measurements. The derived supernova rate is in 
good agreement with the ${\rm SNR}=0.63\,{\rm SNu}$ for irregular galaxies, 
taken from Cappellaro et al. (1997).

\section{Summary}

1. SN\,1995ah is the first supernova observed in a BCD galaxy. It was discovered
during a survey for emission-line galaxies, around 10 days after the maximum
brightness.

2. SN\,1995ah is a Type II supernova and could belong to the rare 
Bright SNII Linear subclass.
We made an attempt to compare our photometric measurement with
the light curve of another Bright SNIIL, SN\,1990K and we found a quite good
agreement, suggesting similar amount of energy involved in the two
explosions. The distance of the supernova from the centre of the galaxy is
2.57$^{\prime\prime}$, at a PA=18.4$^{\circ}$. 

3. The host galaxy HS0016+1449 has
an R=17.7 at a redshift z=0.0147 and its morphologic and spectroscopic
characteristics are typical for the HII galaxy class.

4. From \hspace{-0.08cm} the \hspace{-0.08cm} H${\alpha}$ luminosity of the host 
galaxy, 
 L(H${\alpha}$)=$4.74\times 10^{39}\,$ergs/s, we estimate a 
${\rm SNR} = 0.58\pm0.05\, {\rm SNu}$, in agreement with the SNR for irregular 
galaxies.

\acknowledgements 
We thank to the anonymous referee for useful comments and suggestions.
U. Hopp acknowledges the support by the SFB 375 of the Deutsche 
Fortschungsgemeinschaft.

\end{document}

%% file: eps.tex
\def\eps@scaling{.95}
\def\epsscale#1{\gdef\eps@scaling{#1}}

\def\plotone#1{\centering \leavevmode
    \epsfxsize=\eps@scaling\columnwidth \epsfbox{#1}}

\def\plotfiddle#1#2#3#4#5#6#7{\centering \leavevmode
    \vbox to#2{\rule{0pt}{#2}}
    \includegraphics{#1}}